# The Ethics of Social Media Analytics in Migration Studies

Jamie Mahoney, Northumbria University, jamie.mahoney@northumbria.ac.uk

Kahina Le Louvier, Northumbria University, kahina.le.louvier@northumbria.ac.uk

Shaun Lawson, Northumbria University, shaun.lawson@northumbria.ac.uk

Tel: 01912273388

## Abstract

The prevalence of social media platforms and their use across the globe makes them attractive options for studying large groups of people, particularly when some of these platforms provide access to large amounts of structured data. However, with the collection, storage, and use of this data comes ethical and legal responsibilities, which are particularly important when looking at social groups such as migrants, who are often stigmatised and criminalised.

Various guidelines, frameworks and laws have been developed to ensure social media data is used in the most ethical way. However, they have quickly evolved within the past few years and are scattered across various fields and domains.

To help researchers navigate these issues, this chapter provides an overview of the ethical considerations of studying migration via social media platforms. Building on relevant academic literature, as well as national and supranational frameworks and legislations, we review how the main ethical issues related to social media research have been discussed in the past twenty years and outline good practice examples to mitigate them. This overview is designed to provide researchers with theoretical and practical tools to consider and mitigate the ethical challenges related to social media research in migration-related contexts.

**Introduction**

The prevalence of social media platforms, and their seemingly global use often makes them attractive options for the study of large groups of users or widespread phenomena, events, or topics. In particular, social media platforms provide useful data sources to investigate such a complex and multifaceted phenomenon as migration, for which statistical data is often limited and inconsistent across countries (Migration Data Portal 2020). Social media platforms have been used to study a variety of migration-related aspects, including measuring, monitoring and forecasting migration movements (e.g., Alexander et al. 2020; Maazoli et al. 2020; Martin et al. 2020; Zagheni et al. 2014), detecting hate speech towards migrant populations (e.g., Florio et al. 2019), analysing public opinion and political debate on migration policies (e.g., Siapera et al. 2018), and estimating levels of integration (e.g., Dubois et al. 2018). As such, social media platforms can allow researchers to gather unique insights into migration phenomena that can be used to inform relevant policies and actions.

However, the use of social media data, and its potential for misuse, has resulted in several high-profile cases in the media, such as the Cambridge Analytica scandal in early 2018[1], which have raised a number of crucial ethical issues around the areas of informed consent, privacy, and profiling of individuals. These ethical considerations are particularly important when looking at a population such as migrants, which is often in a vulnerable position in society. Individuals who immigrate to a new country but do not hold the nationality from that country can face risks such as social exclusion (restricted or lack of access to work, housing, social welfare, education, bank accounts, etc), removal from their family, detention, and deportation (Gill 2016; Tazzioli 2020; Yuval-Davis et al. 2018). The disclosure of one's migration status can also lead to stigmatisation and hate speech (ENAR 2018; European Union Agency for Fundamental Rights 2016). As border regimes and immigration control have continuously intensified in European countries and beyond, governments and law enforcement agencies are increasingly relying on big data and social media data to control mobile populations (Latonero and Kift 2018; Metcalfe and Dencik 2019). Individuals who have entered or remained in a country in an irregular way, including people seeking asylum and survivors of trafficking, are particularly at risk, as data published and collected on social media can be used against them (Dimitriadi 2021). In Norway for instance, the police have asked asylum claimants to provide Facebook log-in details to examine the veracity of their accounts and inform their asylum decision (Brekke and Staver 2019).

---

[1] As part of the Cambridge Analytica "scandal" in early 2018, the personal data of approximately 87 million users was accessed without their knowledge and used to create targeted political campaigns. See https://www.bbc.co.uk/news/topics/c81zyn0888lt/facebook-cambridge-analytica-scandal.

In this context, researchers need to pay special attention to the potential harm that collecting and analysing migration-related data from social media platforms can create. Yet, guidance on the ethics of migration research primarily focuses on qualitative methods such as interviews and do not readily apply to social media analytics. Various guidelines, frameworks and laws have been developed to ensure social media data is used in the most ethical way. However, the rapid evolution of social media platforms' terms and conditions, associated privacy policies, and national and supranational legislation make them difficult to navigate.

To draw these different considerations together and help researchers better understand and evaluate the ethical implications of using social media data in migration research, this chapter provides an overview of the main ethical concerns that have been raised in academic literature and other guidelines, and how these may be addressed and mitigated. This chapter especially speaks to researchers who intend to use social media analytics in support of migration and for whom the ethics of this type of research is a new or under-explored topic — whether students, early-career scholars, or primarily qualitative researchers.

This chapter is structured as follows: first, the changing landscape of social media research is explored, outlining how approaches and concerns within academic research and ethical guidelines have changed over time. Following this, these academic, legal, and ethical considerations are then drawn together, outlining some of the main areas of concern, and how these can be addressed and mitigated in practical ways.

## Social media research: A continually changing landscape

Academic research has considered online behaviour in a vast array of contexts, first with the widespread adoption and use of the Internet, and now more recently "Web 2.0" (Di Nucci 1999) and the use of a wide range of social media platforms. The development of this technology and of the research associated with its use has brought up new questions regarding what constitutes an ethical breach and what is considered best practice in this context. In this section, we outline various academic studies and legislations from the late 1990s up to 2021, highlighting the common focuses and concerns, and how these have changed and developed during this period. In doing so, we emphasize that social media research should not be considered static but rather dynamic, with concerns and considerations that arise as new platforms, uses, and data becomes available (or unavailable).

### *Infringing the privacy of individuals*

Prior to the launch of major social media platforms, research related to the use of the Internet and online services highlighted that the amount of data relating to individuals that was becoming available was a major concern (Hudson and Bruckman 2004; Nissenbaum 1998). In the late 1990s researchers were highlighting that the terms "public" and "private" do not apply as easily in an online context as they do in an offline context (Nissenbaum 1998). This is a debate that continues to the present day, with the discussion further complicated by social media users potentially inadvertently making data public, rather than private, and spaces being considered private by some, yet public by others. When considering the ethics of chatroom-based research, Hudson and Bruckman (2004) found that many participants considered online chatrooms to be private spaces, though many are publicly accessible. This reinforces the point made by Bruckman (2004), whereby the notion of "published" or "unpublished" content (taken from more traditional ethical frameworks) is not easily determined with online content, thus further confusing the matter with what might be considered public information in such contexts.

Researchers such as Nissenbaum also highlighted how the adoption of technology and increasing use of networked services would facilitate the large-scale collection and storage of data relating to identifiable individuals, which in turn would raise more questions regarding privacy in these contexts (Nissenbaum 2004). Not only was this data being collected, but as many people are willingly sharing this information online, they are essentially contributing to the potential future violation of their own privacy (Nissenbaum 1998).

While research in more traditional settings often made it easier to determine if information was considered public or private, published or unpublished, these new contexts and the scale of data available presented new ethical challenges for which existing research ethics frameworks and guidelines to online research could not apply. Definitions and safeguards had to be developed and refined around issues such as informed consent, anonymity and profiling, which are particularly sensitive when using social media data in a migration context.

## *Networked publics, misuse of data and informed consent*

Following the launch of major online social networking platforms, such as MySpace, Facebook, YouTube, and Twitter between 2003 and 2006, the focus of academic research and concerns raised within this research began to shift. With the increasing use of these networks, and the scale and scope of data becoming available, there developed an increased focus on the notions of online networks, networked publics, and how the concept of informed consent might be applied, at scale, in these contexts (boyd 2007; Hudson and Bruckman 2004; Fiesler and Proferes 2018; Nissenbaum, 2004).

boyd (2007) highlights four fundamental properties that differentiate unmediated publics from networked publics: persistence, searchability, replicability and "invisible" audiences. These four properties, particularly when combined, demonstrate how data produced, stored, and shared through online social networks presents new ethical considerations and challenges. Not only does the size of the potential audience increase, but the nature of the audience changes, becoming invisible and unknown (Marwick and boyd 2011), and the contexts in which the data can be used – including research studies – is often unknown at the time of posting, with data being retained and readily available, potentially indefinitely. While the original poster may have been aware of the potential audience when they originally shared the content, there is no way to be sure that the poster is aware that the content can still be available, exactly who can access the content, and what it may be used for. This can have severe implications for individuals in vulnerable migration situations where, for instance, a picture posted on social media for friends and family may be used by anti-migration groups to feed racist campaigns (Dearden 2015). Even when anonymised and aggregated as part of a research project on migration flows, the use of geolocated social media data for purposes such as border control and immigration enforcement may not have been anticipated or desired by the original poster. In this context, it is crucial for researchers to consider whether social media users would expect their data to be used in that way, and whether any use of this data could potentially lead to harm.

More recent academic literature has continued to highlight the ongoing questions around informed consent in large-scale social media studies, and started looking at how social media users (as research participants, in this context) perceive the use of their data in research studies (Hudson and Bruckman 2004; Fiesler and Proferes 2018). In a research context, consent means "any freely given, specific, informed and unambiguous indication of the data subject's wishes by which he or she, by a statement or by a clear affirmative action, signifies agreement to the processing of personal data relating to him or her" (EU GDPR, Article 4(11)). Informed consent is a key aspect of human-subject research projects, which also involves the right for participants to withdraw from studies without any consequences to their future treatment. Recognising the importance of informed consent procedures in migration research, as well as the need to address the language and cultural differences that may arise in such research context, have led to publishing various strategies and recommendations (e.g., oral consent, iterative consent, cultural insider) to ensure that participants with a migration background fully understand the implications of being involved in a research project (European Commission 2020; Mackenzie et al. 2007). While these strategies apply to traditional research settings, exactly how, and to what extent, it is possible to gain the informed consent from social media users, continues to be a point of debate in ethical guidelines and academic literature.

Various regulations now indicate that gaining the informed consent of thousands, if not millions, of social media users is impossible and may therefore not always be required for purposes such as research (Association of Internet Researchers 2012; British Sociological Association 2017; General Data Protection Regulation 2016). There is also a shared view that informed consent is not required where the intention to make content public is clear, such as publicly available tweets (Townsend and Wallace 2016). This view, however, is countered somewhat by other research (Fiesler and Proferes 2018) that highlights that many social media users are unaware that their data is being made available to researchers and is being used in research studies without them being consulted or notified. While in a traditional interview setting participants can decide what information they agree to disclose and privacy concerns can be discussed and negotiated, this is not possible in social media analytics studies where users do not directly consent for their data to be used for a specific study. This means that it is the responsibility of the researcher to take all the precautions necessary to identify and minimise any risk of harm.

### *Data Anonymisation and profiling*

As larger and larger datasets became available through social media platforms, examples of the misuse of data, and the potential implications when data can be de-anonymised became more prevalent. Many papers written after 2008 cite the "Tastes, Ties, and Time" project as one such example (Zimmer 2010). This project made use of Facebook profile data of approximately 1,700 students – an entire cohort – from what was intended to be an anonymous US university. In this project, despite reassurances that all identifying information had been removed, it became possible to quickly identify the institution at which the data had been collected, with some degree of certainty, based on seemingly innocuous information that was shared, such as the types of courses offered by the institution and its approximate geographic location. While it was not possible to definitively identify individuals within the dataset, the privacy of the individuals was jeopardised, given the relatively small population with many unique individuals.

This example, and others like it, demonstrates that it is important to not only consider the anonymisation of individual data points and participants, but to also consider how aspects such as data collection, and the broader context of the study is reported. The re-identification of the source of data collection in the "Tastes, Ties, and Time" project was possible through the combination of seemingly innocuous facts that had been reported and has jeopardised the privacy of research participants. A similar situation in a migration-related study could have more serious implications, particularly if individuals are identifiable. For example, this could have revealed individuals' names and locations, their network of family and friends, as well as other sensitive information including immigration status. This could potentially expose individuals to harm such as hate speech, detention or, in the case of refugees, pressures from authorities on family members who remained in the country of origin (Bloemraad and Menjívar 2021).

While in the "Tastes, Ties, and Time" project, the re-identification was not intended, other studies and organisations have purposefully used social media data for profiling, a technique that consists in using automated means to categorise individuals according to their personal characteristics (EU GDPR, Article 4(2)). For instance, Jung et al. (2017) apply a facial recognition algorithm to social media account profile pictures to infer attributes such as gender, race, and age and use this demographic information to inform audience segmentation. As well as targeted advertising, social media profiling has been used for behaviour prediction, including migration behaviour (Zagheni et al. 2017). Such automated detection and profiling of personal data without explicit consent, including special category data such as racial or ethnic origin, can expose individuals to a surveillance that can lead to social exclusion, prejudice and discrimination (Mitrou et al. 2014). The use of these techniques by law enforcement agencies in Europe to detect and prevent migration arrivals may also have disruptive effects that force individuals into new and more dangerous migration routes (Dimitriadi 2021).

In light of these problematic uses of social media profiling, and to respond to changes in legislation, including the EU GDPR and associated national laws, many online platforms and services have altered how they had previously operated, thus impacting how research studies are planned and conducted. For example, Facebook and Instagram have now further restricted access to their previously public Application Programming Interface (API) endpoints, which provided easy access to public data for researchers and other individuals. As a result of these operational changes, much data that had previously been available to researchers became further restricted, or entirely unavailable – in some cases with little or no warning. While more recent studies that consider migration have used Instagram as a data source (Jaramillo-Dent and Pérez-Rodríguez 2019), these studies often now focus on smaller collated datasets, as opposed to the larger-scale datasets that could previously be collected using automated tools and the public APIs.

This overview of the main ethical concerns over the use of social media shows that progress has been made in the way we understand the risks linked to this type of research. Guidelines and legislations have been introduced to further protect individual data privacy, and social media have had to enhance control over the kind of data that can be processed through their platforms. However, this overview also shows that both the social media and political landscapes are constantly changing, leading to new risks being created for populations such as migrants, who are increasingly subjected to digital surveillance, social exclusion and stigmatisation. It is therefore important for researchers engaging in social media research in a migration context to consider these ethical issues carefully. In the next section, we outline key legislations, tools and principles that can help researchers to develop ethical research programmes.

## Common Areas of Concern and Potential Mitigation Measures

In this section, we draw together the previous discussions highlighting common areas of concern when designing and conducting social media-based research. By drawing on relevant legislations, as well as examples of guidelines published by professional bodies, we provide practical recommendations for how such concerns may be mitigated.

## *Observing public behaviour in public situations*

Many of the concerns raised in the previously explored academic literature focus on participants' and researchers' views on the public or private nature of aspects of social media platforms, particularly in relation to its impact on the need for informed consent from social media users.

Guidelines, such as those published by the British Psychological Society, recommend that "observation of public behaviour needs to take place only in public situations where those observed would expect to be observed by strangers" (British Psychological Society 2021, p.8). Applied to a social media context, this would suggest that only content that is clearly public, such as public tweets or Instagram posts, should be collected as part of any study of social media content. The use of public APIs provided by these platforms, if they are made available, helps in this regard, as these provide clear levels of access, with only public information (such as tweets) being accessible without the user's express permission.

In situations where the public nature of the data is not clear – perhaps in semi-public environments, such as discussion groups on platforms such as Facebook, researchers should take into account any potential damaging effects on participants that undisclosed observations may have. As part of making any decision as to whether such data should be used, researchers should also consider if obtaining consent is necessary in this context (British Psychological Society 2021). These considerations may be particularly important in migration-related studies where, for example, individuals may be posting in a discussion group about migration-related experiences. The semi-closed nature of discussion groups may have led the posters to believe that the content was private, and the potential for causing harm to these individuals would be increased if the data was publicised. In such a situation then those conducting the research would have to carefully consider if gaining informed consent was appropriate.

Limiting the collection of data to those areas that can clearly be considered as public means that informed consent is not necessarily required. Under EU GDPR Article 14 Paragraph 5(b) obtaining informed consent from data subjects is not required when "the provision of such information proves impossible or would involve a disproportionate effort, in particular for processing for archiving purposes in the public interest, scientific or historical research purposes or statistical purposes". This applies to social media research where large amounts of data are being collected. However, this should not be considered as the end of researchers' responsibilities to protect the individuals included in these studies. Further safeguards, such as those explored in the following sub-sections, should also be considered, in order to minimise the risks that research participants are exposed to as a result of any research being conducted.

*Maintaining anonymity of research participants*

Maintaining the anonymity of research participants is another key aspect of social media-based research, particularly when working with vulnerable populations and in sensitive contexts, such as migrants and migration-related contexts. Anonymisation refers to the various techniques that can be used to convert personal data into anonymised data (ICO, 2012), that is, data that "does not relate to an identified or identifiable natural person or to personal data rendered anonymous in such a manner that the data subject is not or no longer identifiable" (EU GDPR, Recital 26). The process of anonymising data is designed to protect individual research participants and any information that may be considered personal or private. Anonymisation also satisfies legal requirements, such as EU GDPR.

Generally, it is considered best practice (particularly when it is not possible to gain the original poster's consent) to avoid the verbatim quoting of any individual's social media content in research publications (Fiesler & Proferes 2018). This is because many platforms have search functionality, and therefore the content of a post could be used to identify the original poster using these functions. In many cases, quotes that may be used as examples are often paraphrased or merged (British Psychological Society 2021). In doing so, the original context and meaning can be preserved, while making it more difficult for the original authors to be identified. Further, where accounts belonging to individuals are mentioned within these posts, they are often replaced with generic terms, such as @user1, for example.

While such processes do not fully remove the possibility of individuals being identifiable based on reports and publications, it does reduce the risk, which is an important consideration, particularly when seeking to protect the anonymity and potentially vulnerable research participants. Other steps, including minimising the amount of data being collected and stored, and anonymising or pseudonymising the collected data at the earliest opportunity, can also reduce the level of risk that participants may be exposed to as the result of any potential data breaches. Pseudonymisation is "'pseudonymisation' means the processing of personal data in such a manner that the personal data can no longer be attributed to a specific data subject without the use of additional information, provided that such additional information is kept separately and is subject to technical and organisational measures to ensure that the personal data are not attributed to an identified or identifiable natural person" (EU GDPR, Article 4(5)). Unlike anonymisation, pseudonymisation does not remove all identifying information from the data but reduces the linkability of a dataset with the original identity of an individual. This method is particularly useful during the data analysis phase, when some form of identification may still be required.

### *Avoiding profiling of participants*

Profiling and automated decision making is covered by EU GDPR, as well as other national legislation such as the UK Data Protection Act (UK Government 2018), and as such the analyses undertaken within research projects, and particularly the effect of these analyses on participants, need to be carefully considered.

Recital 71 of GDPR, for example, states:

> *Such processing includes 'profiling' that consists of any form of automated processing of personal data evaluating the personal aspects relating to a natural person, in particular to analyse or predict aspects concerning the data subject's performance at work, economic situation, health, personal preferences or interests, reliability or behaviour, location or movements, where it produces legal effects concerning him or her or similarly significantly affects him or her.*

Some of these aspects are particularly relevant to migration-related studies and need to be addressed carefully. While some form of profiling is often inherent in social media studies, as a means of understanding who is sharing information on these platforms, careful consideration should be given to where and for what purpose this profiling will be undertaken, or may be possible from the analyses conducted, and should be addressed appropriately when planning and conducting research. Data anonymisation and aggregation, for example, are two ways in which potentially problematic profiling of individuals can be avoided, as it negates the possibility of linking any such profiling or decision making back to an identifiable individual (Townsend and Wallace 2016).

### *Evaluating and mitigating risks*

Throughout the legislations, guidelines, and academic research that has been discussed within this chapter, one of the key principles has been to minimise potential harms to research participants, while maximising the benefits of the research being conducted.

One way in which potential risks can be identified, evaluated, and measures put in place to mitigate these risks is through processes such as conducting a Data Protection Impact Assessment (DPIA). DPIAs are required by Article 35 of EU GDPR, where any processing of data may result in a high risk to the rights and freedoms of individuals. Rather than a static document, these DPIAs should be seen as a live document and process, which should be referred to and updated as the project develops and research activities are conducted.

As migration-related research using social media data will likely involve the collection and use of personal information, the DPIA process forms a key aspect of identifying and reducing the risks involved for those taking part in the research. Such risks may include individuals being identifiable based on how results and findings are reported, the release of raw data, including personally identifiable information, or the potential for inappropriate profiling of individuals or groups based on the collected and analysed data.

The previous sub-sections have discussed some common areas where these risks can be mitigated such as through the collection of public data, maintaining the anonymity of research participants and avoiding the profiling of individuals where possible. While this list is not exhaustive, and each research study will present its own challenges and potential risks, these core principles can be used alongside further guidelines set by organisations such as the Association of Internet Researchers (Association of Internet Researchers 2012; Association of Internet Researchers 2019), the British Psychological Society (British Psychological Society 2021) and the British Sociological Association (British Sociological Association 2017a; British Sociological Association 2017b) to identify and address the specific risks of each individual study.

**Conclusion**

This chapter has provided an overview of ethical and legal considerations related to the use of social media data in research projects, particularly migration-related studies, which are often sensitive in nature, and may involve working with potentially vulnerable populations. An overview of related academic literature highlights not only how the focus of ethical concerns and best practice develops over time, but also how approaches need to remain flexible, and respond to these changes, and changes imposed by social media platforms. Strategies to evaluate and mitigate the risks associated with (the lack of) informed consent, anonymity and profiling have been outlined. These strategies will aid researchers working with social media data in migration studies to evaluate and mitigate those risks in a way which adheres to the various laws and guidelines, resulting in beneficial research which protects the rights of the individuals being studied.

Townsend, L. and Wallace, C. (2016). Social media research: A guide to ethics. Technical Report, The University of Aberdeen. https://www.gla.ac.uk/media/Media_487729_smxx.pdf.

UK Government. (2018). UK Data Protection Act. https://www.legislation.gov.uk/ukpga/2018/12/contents/enacted. Accessed. 24 June 2021.

Yuval-Davis, N., Wemyss, G., & Cassidy, K. (2018). Everyday bordering, belonging and the reorientation of British immigration legislation. *Sociology*, *52*(2), 228-244.

Zagheni, E., Garimella, V. R. K., Weber, I., & State, B. (2014). Inferring international and internal migration patterns from twitter data. *Proceedings of the 23rd International Conference on World Wide Web, WWW '14 Companion*, New York, NY: ACM, 439–444.

Zagheni, E., Weber, I., & Gummadi, K. (2017). Leveraging Facebook's advertising platform to monitor stocks of migrants. *Population and Development Review, 43*(4), 721-734. http://www.jstor.org/stable/26622775. Accessed 18 June 2021.

Zimmer, M. (2010). "But the data is already public" : On the ethics of research in Facebook. *Ethics and Information Technology, 12(4),* 313-325.

## Biographies

Jamie Mahoney is a Research Fellow in the Department of Computer & Information Sciences at Northumbria University. His research interests include the use of social media in various aspects of people's lives, including commerce, health, marketing, and politics.

Kahina Le Louvier is a Research Fellow in the Department of Computer & Information Sciences at Northumbria University. Her research interests include information practices and the effects of asylum and immigration policies.

Shaun Lawson is Professor of Social Computing and Head of the Department of Computer & Information Sciences at Northumbria University. He leads the Northumbria Social Computing (NorSC) research group.

## Funding Acknowledgement

This research is conducted as part of the PERCEPTIONS H2020 project which has received funding from the European Union's H2020 research & innovation programme under Grant Agreement No. 833870.